\documentstyle[seceq,epsf,wrapfig,twoside]{ptptex}
\def\Journal#1#2#3#4{{#1}{\bf #2} (#4), #3}
\def\PTP{Prog.~Theor.~Phys.~}

\def\NPA{Nucl.~Phys. \bf{A}}
\def\NPB{Nucl.~Phys. \bf{B}}
\def\PLB{Phys.~Lett. \bf{B}}

\def\PRL{Phys.~Rev.~Lett.~}
\def\PRD{Phys.~Rev. \bf{D}}
\def\ZPC{Z.~Phys. \bf{C}}

\def\PR{Phys.~Rev.~}

\notypesetlogo  
\title{%
Relation between Scattering and Production Amplitudes\\
---{\em concerning $\sigma$ Particle 
in $\pi\pi$ System}---
}
\author{%
Muneyuki {\sc Ishida}, Shin {\sc Ishida}$^*$ and Taku {\sc Ishida}$^{**}$
}
\inst{%
Department of Physics, University of Tokyo , Tokyo 113, Japan\\
$^*$ Atomic Energy Research Institute, College of Science and Technology\\
Nihon University, Tokyo 101, Japan\\ 
$^{**}$ KEK, Oho, Tsukuba 305, Japan
}
\recdate{%
March 1, 1998
}
\abst{%
In a series of previous papers the present authors and collaborators
have shown strong evidence for existence of the long-sought 
$\sigma$-particle by analyzing the $\pi\pi$-scattering and 
$\pi\pi$-production processes, applying, respectively, the 
IA method and the VMW method.
In this paper 
we examine the relation between the scattering 
amplitude ${\cal T}$ and the production amplitude ${\cal F}$
from the viewpoint of the unitarity and the final state interaction 
(FSI) theorem by using a simple field theoretical model.
As a result it is shown that  the amplitudes in the  physical state 
representation are directly represented through the  Breit-Wigner 
amplitudes of the relevant resonances, and the respective forms of
 ${\cal T}$ and ${\cal F}$ coincide with those in the IA and VMW methods,
justifying our methods of analyses.
}

\begin{document}
\maketitle

\setcounter{tocdepth}{4}

\section{Introduction}
The light iso-singlet scalar $\sigma$ meson appears as a chiral partner 
of the $\pi$ meson in the linear representation of chiral symmetry.
In the Nambu$\cdot$Jona-Lasinio (NJL) model\cite{rf:NJL} and its
extended version (ENJL)\cite{rf:kunih0,rf:enjl,rf:taki,rf:klimt}
${}^{,}$
\footnote{
The ENJL model predicts
the existence of the NG pseudo-scalar boson, scalar meson,
and, moreover, vector and axial-vector meson nonets.}
adapted to the quark model, which simply realizes
the physical situation of D$\chi$SB of QCD, the existence
of the $\sigma$-meson (or scalar meson nonet) is predicted 
with mass$\simeq 2m_q$ ($m_q$ being
the constituent quark mass).
This $\sigma$ gives quarks constituent masses,
and in this sense it partly plays a role of the Higgs particle of QCD.
The existence of the $\sigma$-meson has been long-sought 
from various viewpoints, both theoretically 
and phenomenologically.\cite{rf:NJL,rf:kunih0,rf:enjl,rf:taki,rf:klimt,rf:LsM,rf:kramer,rf:BL,rf:CH,rf:obe,rf:sca,rf:shim,rf:kin,rf:shaba,rf:Bentz,rf:MLS,rf:achasov,rf:rudaz}
However, its existence  as a resonant particle 
has not yet been generally accepted.
A major reason for this is due to the negative results of 
conventional 
analyses\cite{rf:Morgan,rf:ff,rf:morg,rf:MP2,rf:aniso}
${}^{,}$
\footnote{
However, see the analyses\cite{rf:menne,rf:beve,rf:svec,rf:zou}
which suggest the existence of $\sigma$.
 }
of the $\pi\pi$ phase-shift obtained from the high-statistics 
data of a CERN-Munich experiment\cite{rf:CM} in 1974.

In the recent $pp$-central collision experiment,
a huge event concentration in the $I=0$ $S$-wave $\pi\pi$ channel
is seen\cite{rf:GAMS} 
in the region of $m_{\pi\pi}$ around $500\sim 600$ MeV,
which is too large to be explained as a simple
``background" and strongly suggests the  
existence of a resonant particle, which can be 
identified with $\sigma$.
Actually, it has been shown that the characteristic shape 
of the $\pi^0\pi^0$ effective mass spectra below 1 GeV can
 be explained\cite{rf:taku1,rf:taku2,rf:had97d} by  
a coherent sum of the two ($\sigma$ and $f_0$) 
Breit-Wigner resonant amplitudes. This parametrization method of 
the $\pi\pi$-production amplitude is called
the VMW method.\cite{rf:sawa} 

A similar event concentration
around the region of $m_{\pi\pi}=500\sim 600$ MeV 
is also observed\cite{rf:JPsi} in 
the $\pi\pi$ system obtained in the 
 $J/\Psi\rightarrow\omega\pi\pi$ decay,
suggesting the existence of an iso-singlet $S$-wave state,
which also seems to be identified with $\sigma$. 

However, the claim of $\sigma$-existence has been
criticized from the so-called 
``universality of $\pi\pi$-scattering amplitude"
argument as follows:\cite{rf:pennington}
  ``unitarity requires a
resonance that decays to $\pi\pi$, for example,
has to couple in the same way to this final state
whether produced in $\pi\pi$ scattering or centrally
in $pp\rightarrow pp(\pi\pi )$...Thus claims of a 
narrow $\sigma$(500) in the GAMS results cannot be
correct as no such state is seen in $\pi\pi$ scattering."

On the other hand being inspired by these experiments,
we and our collaborators have recently made 
a re-analysis\cite{rf:pipip,rf:further,rf:had97a} of 
the $\pi\pi$ phase shift through a new $S$-matrix parametrization
method, the interferring amplitude(IA) method,
and found  strong evidence 
for the existence of the $\sigma$ particle.
The reason we obtained a different result from that in 
the conventional analyses is due to the introduction of the 
repulsive background phase $\delta_{\rm BG}$, which cancels a main
part of the attractive phase due to $\sigma$ production.
This cancellation mechanism is guaranteed by 
chiral symmetry.\cite{rf:MY,rf:had97b} However, 
it has been overlooked in the 
conventional analyses,\cite{rf:morg,rf:MP2} leading to the wrong
conclusion against the $\sigma$ existence.
Several other groups have independently performed
re-analyses also leading to a positive 
conclusion\cite{rf:achasov,rf:kamin,rf:torn,rf:hara} 
for $\sigma$ existence.\footnote{
Reflecting these results
the $\sigma$-particle has been revived in the list of
the latest edition of PDG,\cite{rf:pdg96}
after missing for twenty years,
with the somewhat tentative label,`` $f_0(400\sim 1200)$
or $\sigma$".
}
Thus, not only the above ``universality''argument 
for the $\pi\pi$-production processes but also the contents
of the ``universality'' itself 
must be re-considered.

In this paper
we examine\cite{rf:had97,rf:had97c} the validity of the methods,
the IA method (VMW method) for
scattering (production) amplitude
applied in the above phenomenological analyses 
leading to $\sigma$ existence, 
from the general viewpoint of the unitarity 
and the applicability of 
FSI-theorem, especially noting the relation to
the universality argument. 


\section{General problem}

In treating the $\pi\pi$-scattering and production amplitudes,
there are two general problems to be taken into account.

$\underline{Unitarity}$\ \ \  The scattering 
amplitude ${\cal T}$ (and its
hermitian conjugate ${\cal T}^\dagger$) must satisfy
the condition
\begin{eqnarray}
{\cal T}-{\cal T}^\dagger &=& 2i{\cal T}\rho{\cal T}^\dagger ,
\label{eq:uniT}
\end{eqnarray}
where $\rho$ is the $\pi\pi$-state density.

$\underline{ Final\ State\ Interaction(FSI)\ theorem}$\ \ \  The 
production amplitude ${\cal F}$ must have the same 
phase\cite{rf:watson} as  ${\cal T}$,
\begin{eqnarray}
{\cal T}\propto e^{i\delta} &\rightarrow& 
{\cal F}\propto e^{i\delta},
\label{eq:FSI}
\end{eqnarray}
in case that the initial states have 
no strong phases.\footnote{ See 
detailed discussion in \S 3.4.} \\
    \\
({\em ``Universality" of scattering amplitudes})\ \ \ 
Furthermore, a more restrictive relation
between  ${\cal F}$ and  ${\cal T}$
is conventionally required on the basis of the 
``Universality of ${\cal T}$,"\cite{rf:pennington,rf:morg}  
as mentioned in Introduction: That is, 
\begin{eqnarray}
{\cal F} &=& \alpha (s){\cal T},
\label{eq:FaT}
\end{eqnarray}
with a smooth real function $\alpha (s)$ of $s$.
Following this, most 
analyses of experimental data on ${\cal F }$
in the region $m_{\pi\pi}\stackrel{<}{\scriptstyle \sim} 1.5$ GeV,
obtained in any type of production processes,
are made as follows.
First, ${\cal T}$, with comparatively more rigorous
data, is analyzed. Then, using this result as input,
${\cal F}$ is analyzed by parametrizing the 
function $\alpha (s)$.
Thus, in this procedure, physical information
can be obtained only through scattering experiments,
while any production experiment loses its value
in seeking new resonances.

Our results obtained by phenomenological
analysis, using the VMW-method, 
 on the production process 
 was criticized\cite{rf:pennington}  
along the above line of thought that the claims for $\sigma$ in
${\cal F}$ cannot be correct, as no such resonant
poles exist in ${\cal T}$ due to the conventional 
phase shift analyses. However, as a result of the re-analysis
of the $\pi\pi$ scattering with the IA method, which satisfies
unitarity, there seems to indeed exist a $\sigma$ pole 
in ${\cal T}$. Accordingly, the main reason for the above 
criticism has been lost. 
However, in the VMW method there still remains a problem; 
whether or not it is consistent with
the FSI theorem.

In the following we re-examine the relation
between ${\cal F}$ and ${\cal T}$ 
concretely, by using a simple model.\cite{rf:aitchson,rf:chung}

\begin{figure}[t]
 \epsfysize=3. cm
 \centerline{\epsffile{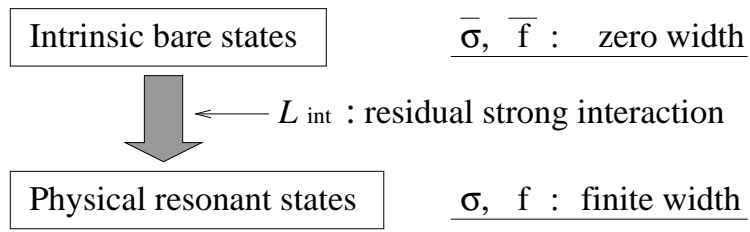}}
 \caption{Resonance view based 
on a simple field-theoretical model.}
\label{fig:model}
\end{figure}

\noindent ({\em Simple field theoretical model})\ \ \ \ 
In the NJL-type model as a low energy effective theory of QCD,
(and in the linear $\sigma$ model, L$\sigma$M, 
obtained as its local limit),
or 
in the constituent quark model, 
the pion $\pi$ and
the resonant particles such as $\sigma$(600) or $f_0(980)$
are the color-singlet $q\bar q$-bound states and  
are treated equivalently.
These  
``intrinsic quark dynamics states,"
denoted as 
$\bar \pi,\ \bar \sigma $ and $\bar f$
are stable particles with zero widths and appear from the beginning. 
Actually, these particles 
have structures 
and interact with one another 
(and a production channel ``$P$'') 
through the residual strong interaction
\begin{eqnarray}
{\cal L}^{\rm scatt}_{\rm int} = \bar g_\sigma\bar\sigma\pi\pi
+\bar g_f\bar f\pi\pi+\bar g_{2\pi}(\pi\pi )^2
({\cal L}^{\rm prod}_{\rm int} = \bar \xi_\sigma\bar\sigma ``P"
\hspace{-0.1cm}+\bar \xi_f\bar f``P"\hspace{-0.1cm}+
\bar\xi_{2\pi}(\pi\pi ) ``P").\ \ \ \ \ \ \ \ \ \ 
\label{eq:Lint}
\end{eqnarray}
Due to this,
these bare states change\cite{rf:achasov,rf:torn} into
physical states, denoted by
$\pi (=\bar\pi ),\ \sigma$ and $f$ with finite widths,
as shown in Fig. \ref{fig:model}.
In the following we consider only the
virtual two $\pi$-meson effects 
for the resonant $\sigma$ and $f$ particles.

\section{Three Different ways of description of scattering amplitudes}

\begin{figure}
 \epsfysize=15. cm
 \centerline{\epsffile{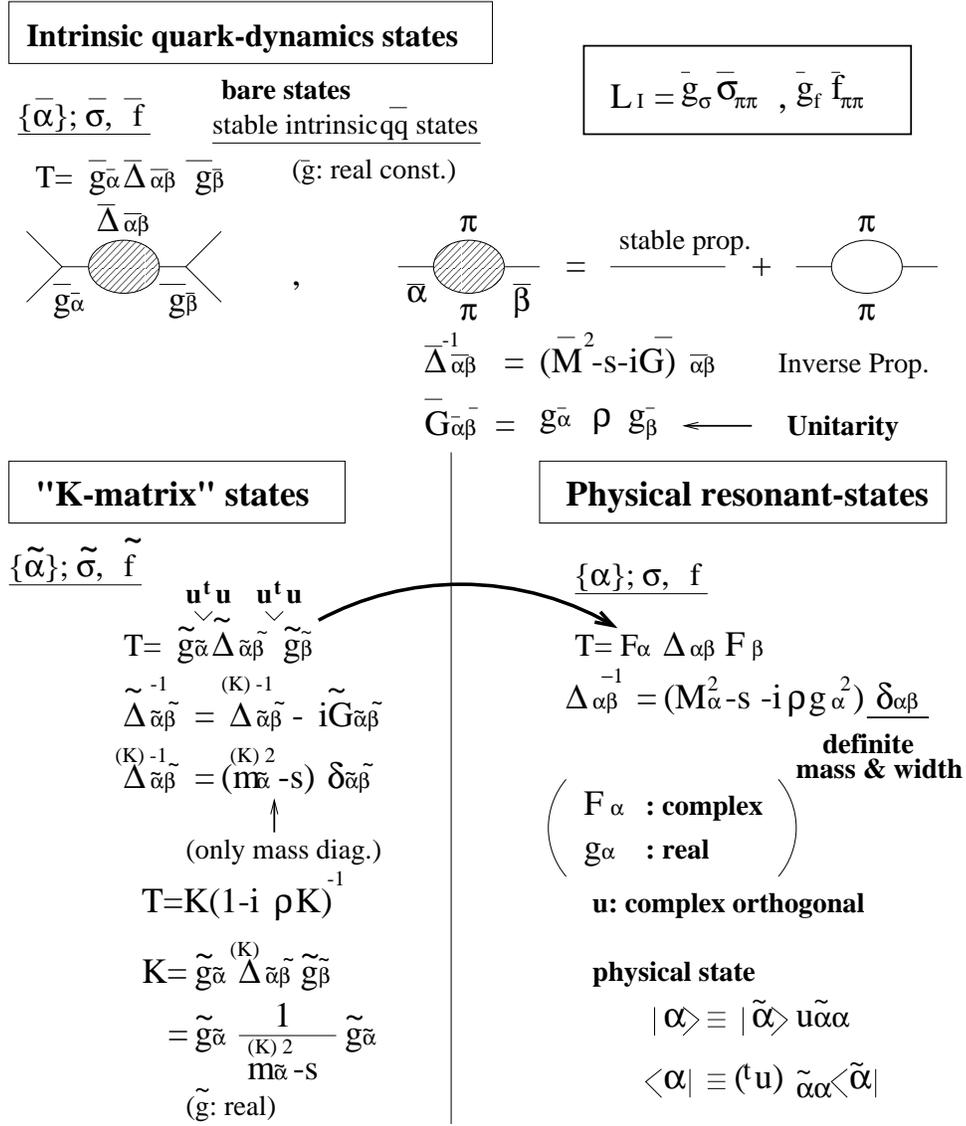}}
 \caption{ Three different representations of 
scattering amplitudes. }
\label{fig:ways}
\end{figure}

There are the following three ways to represent
 scattering amplitudes,
corresponding to the three types of basic 
states for describing the resonant particles,
as is depicted in Fig. \ref{fig:ways}.

\begin{description}

\item[1.\ ] {\it Intrinsic\ quark-dynamics\ states\ (bare states)
representation}\\
In the bases of zero-width bare states,
denoted as $|\bar \alpha\rangle$,
the $\pi\pi$-scattering amplitude is represented in terms of 
the $\pi\pi$-coupling constants $\bar g_{\bar\alpha}$
 and the propagator matrix $\bar\Delta$ as
\begin{eqnarray}
{\cal T} &=& \bar g_{\bar\alpha}\bar\Delta_{\bar\alpha\bar\beta}
\bar g_{\bar\beta}.
\label{eq:TinB}
\end{eqnarray}
By taking into account the effects of repetition of 
the $\pi\pi$-loop,
the bare-state propagator
acquires a finite width.
The corresponding
inverse propagator is represented by
\begin{eqnarray}
\bar\Delta_{\bar\alpha\bar\beta}^{-1}
 &=& (\bar M^2-s-i\bar G)_{\bar\alpha\bar\beta}.
\label{eq:DinB}
\end{eqnarray}
The real and imaginary parts of the squared mass matrix
take non-diagonal forms, which implies that 
the bare states have indefinite masses and lifetimes.
The imaginary part of the inverse propagator is 
\begin{eqnarray}
\bar G_{\bar\alpha\bar\beta}
 &=& \bar g_{\bar\alpha}\rho\bar g_{\bar\beta};\ \ 
(\rho =\sqrt{1-4m_\pi^2/s}/16\pi ),
\label{eq:Gbar}
\end{eqnarray}
where $\rho$ is the $\pi\pi$-state density. 
Then our ${\cal T}$ is easily shown to satisfy 
the unitarity, Eq. (\ref{eq:uniT}).

\item[2.\ ] ``${\cal K}$-{\it matrix"\ states representation}\\
The real part of $\bar\Delta^{-1}$ is symmetric and
can be diagonalized by an orthogonal transformation:
It transforms the bare states 
$|\bar\alpha\rangle$ into the
``${\cal K}$-matrix" states 
 $|\tilde \alpha\rangle$, as
\begin{eqnarray}
|\tilde\alpha\rangle &\equiv& |\bar\alpha\rangle 
o_{\bar\alpha\tilde\alpha}.
\label{eq:diagK}
\end{eqnarray}
Correspondingly,
${\cal T}$ is represented by
\begin{eqnarray}
{\cal T} &=& \tilde g_{\tilde\alpha}
\tilde\Delta_{\tilde\alpha\tilde\beta}
\tilde g_{\tilde\beta},
\label{eq:TinK}
\end{eqnarray}
with the inverse propagator
\begin{eqnarray}
\tilde\Delta_{\tilde\alpha\tilde\beta}^{-1}
=(\Delta^{-1}_{{\cal K}}-i\tilde G)_{\tilde\alpha\tilde\beta}
 &;&
\ \ \ \ {\Delta^{-1}_{{\cal K}}}_{\tilde\alpha\tilde\beta}
= (\tilde m^2_{\tilde\alpha}-s)
\delta_{\tilde\alpha\tilde\beta},\ \ \ \ 
\tilde G_{\tilde\alpha\tilde\beta}=\tilde g_{\tilde\alpha}
\rho\tilde g_{\tilde\beta},
\label{eq:DinK}
\end{eqnarray}
where the coupling constant
 $\tilde g_{\tilde\alpha}
(=\bar g_{\bar\alpha}o_{\bar\alpha\tilde\alpha})$
is real.
These states have definite masses but indefinite lifetimes.
The propagator $\tilde\Delta$ is, owing to 
Eq. (\ref{eq:DinK}), expressed in the form
representing concretely 
the repetition of the $\pi\pi$ loop, 
as 
\begin{eqnarray}
\tilde\Delta =(1-i\Delta_{{\cal K}}\tilde G
)^{-1}\Delta_{{\cal K}}
&=& \Delta_{{\cal K}}+i\tilde\Delta\tilde G
\Delta_{{\cal K}}.
\label{eq:DinK2}
\end{eqnarray}
Then ${\cal T}$, similarly as  
the ${\cal K}$ matrix representation in potential theory,  
takes the form
\begin{eqnarray}
{\cal T} &=& {\cal K}+i{\cal T}\rho{\cal K}
= {\cal K}(1-i\rho{\cal K})^{-1};\ \  
 {\cal K} = \tilde g_{\tilde\alpha}
\Delta_{{\cal K}\tilde\alpha\tilde\beta}
\tilde g_{\tilde\beta}
=\tilde g_{\tilde\alpha}
(\tilde m_{\tilde\alpha}^2-s)^{-1}
\tilde g_{\tilde\alpha}.\ \ \ \ \ \ \ \ \ 
\label{eq:Krep}
\end{eqnarray}
From the viewpoint of the present 
field-theoretical model, 
this ``${\cal K}$ matrix,"
Eq. (\ref{eq:Krep}), has the physical meaning
of the  
propagators of bare particles
with infinitesimal imaginary widths,
$\tilde m_{\tilde\alpha}^2\rightarrow
\tilde m_{\tilde\alpha}^2-i\epsilon$,
while the original 
${\cal K}$ matrix in potential theory is purely real
and has no direct meaning.

\item[3.\ ] {\it Physical\ resonant\ states representation},\\
The imaginary part of 
$\tilde\Delta^{-1}$
in the ${\cal K}$-matrix state representation
was remained in a non-diagonal form.
$\tilde\Delta^{-1}$ can be diagonalized by 
a complex orthogonal\cite{rf:aitchson,rf:rosen} matrix $u$,
satisfying ${}^tuu=1$.
It transforms 
$|\tilde\alpha\rangle$
into the unstable physical states $|\alpha\rangle$ as  
\begin{eqnarray}
|\alpha\rangle &\equiv& 
|\tilde\alpha\rangle u_{\tilde\alpha\alpha},\ \ \ \ 
\langle\alpha |\equiv^tu_{\alpha\tilde\alpha} 
\langle\tilde\alpha | .
\label{eq:diag}
\end{eqnarray}
It is to be noted that
the transformation is not unitary and 
$\langle\alpha |\neq (|\alpha\rangle )^\dagger$.
Correspondingly, 
the ${\cal T}$ matrix is represented by
\begin{eqnarray}
{\cal T} &=& F_{\alpha}\Delta_{\alpha\beta}F_{\beta}
=\sum_\alpha 
F_\alpha (\lambda_\alpha -s)^{-1}F_\alpha ;\ \ \ 
F_{\alpha}(\equiv \tilde g_{\tilde\beta}u_{\tilde\beta\alpha})
\label{eq:TinP}
\end{eqnarray}
where $\lambda_\alpha$ is the physical squared mass of the
$|\alpha\rangle$ state, and  the 
$F_{\alpha}$
are the physical coupling constants, which are generally complex.
The physical state has a 
definite mass and lifetime,
and is observed as a resonant particle
directly in experiments.

\end{description}

\section{Scattering and production amplitudes 
consistent to FSI-condition}

In the following  
we show how the formulas in the IA and VMW methods 
satisfying the FSI theorem are derived effectively 
in the physical state representation. 
We start from the ``${\cal K}$ matrix" states,
which can be identified
with the bare states
$|\bar \alpha\rangle (\equiv |\tilde \alpha\rangle )$
without loss of essential points, 
since the reality
of the coupling constant is unchanged
through the orthogonal transformation Eq. (\ref{eq:diagK}).
The real part of the mass correction generally does not
have a sharp $s$ dependence, and, accordingly, 
 $\tilde g$ is 
almost $s$ independent, except for in  
the threshold region.

\subsection{Derivation of 
IA-metod and VMW-method satisfying FSI-theorem}

First we consider the two ($\bar\sigma ,\ \bar f$) 
resonance-dominating case,
assuming $\bar g_{2\pi}$\\
$=\bar\xi_{2\pi}=0$.
The scattering amplitude ${\cal T}$ 
in the bare state representation 
is given by Eq. (\ref{eq:Krep}) as
\begin{eqnarray}
{\cal T}^{\rm Res} 
  = 
{\cal K}^{\rm Res}/(1-i\rho{\cal K}^{\rm Res});\ \ 
{\cal K}^{\rm Res}
=\bar g_{\bar\sigma}(\bar m_{\bar\sigma}^2-s)^{-1}\bar g_{\bar\sigma}
+\bar g_{\bar f}(\bar m_{\bar f}^2-s)^{-1}\bar g_{\bar f}.
\ \ \ \ \ \ \ \ \ \ \ \ \ \ \ \ \ \ \ \ \ \ \ \ 
\label{eq:Kresrep}
\end{eqnarray}
\begin{figure}[t]
 \epsfysize=3. cm
 \centerline{\epsffile{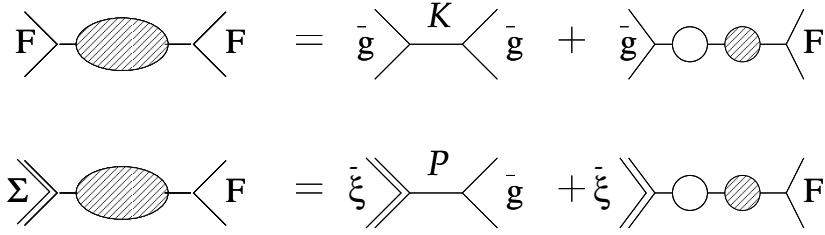}}
 \caption{
Scattering and production mechanism in 
a simple field-theoretical model 
of resonance dominative case.
The production amplitude is obtained, 
following the mechanism 
shown in the figure,
 by replacing the 
first $\pi\pi$-coupling 
constant $\bar g$ in ${\cal T}$
with the production coupling $\bar\xi$.  
The  ${\cal F}$ amplitude obtained in this 
way automatically satisfies the 
FSI theorem. }
\label{fig:prod}
\end{figure}
The production amplitude ${\cal F}$ is obtained
by replacing  $\bar g^2$, appearing in the numerator
${\cal K}$ of ${\cal T}$, by $\bar g \bar\xi$ as
\begin{eqnarray}
{\cal F}^{\rm Res}
 = {\cal P}^{\rm Res}/(1-i\rho{\cal K}^{\rm Res});\ \ 
{\cal P}^{\rm Res}
=\bar\xi_{\bar\sigma}(\bar m_{\bar\sigma}^2-s)^{-1}\bar g_{\bar\sigma}
+\bar\xi_{\bar f}(\bar m_{\bar f}^2-s)^{-1}\bar g_{\bar f},
\ \ \ \ \ \ \ \ \ \ \ \ \ \ \ \ \ \ \ \ \ \ \ \ 
\label{eq:Presrep}
\end{eqnarray}
where ${\cal P}^{\rm Res}$ is 
the production ``${\cal K}$ matrix".
The FSI theorem is automatically satisfied,
since both ${\cal K}^{\rm Res}$ and ${\cal P}^{\rm Res}$
can be treated as real and
the phases of ${\cal T}^{\rm Res}$ and ${\cal F}^{\rm Res}$
come from the common factor $(1-i\rho{\cal K}^{\rm Res})^{-1}$.

In the
physical state representation, 
 ${\cal T}$ is given by Eq. (\ref{eq:TinP}) as
\begin{eqnarray}
{\cal T}^{\rm Res} &=& 
F_\sigma (\lambda_\sigma-s)^{-1}F_\sigma   
+F_f(\lambda_f-s)^{-1}F_f,
 \label{eq:TresinP}
\end{eqnarray}
where the physical squared mass $\lambda_\alpha$
is given by 
\begin{eqnarray}
\lambda_\alpha &=& (1/2)[\bar m_{\bar\sigma}^2+\bar m_{\bar f}^2
-i\rho (\bar g_{\bar\sigma}^2+\bar g_{\bar f}^2)
\pm \sqrt{(\bar m_{\bar\sigma}^2-\bar m_{\bar f}^2
-i\rho (\bar g_{\bar\sigma}^2-\bar g_{\bar f}^2))^2
-4\rho^2\bar g_{\bar\sigma}^2\bar g_{\bar f}^2 }];\nonumber\\ 
  &\equiv& M_\alpha^2-i\rho g_\alpha^2.\ \ \ \ \ \ \ \ 
\alpha =f,\sigma .
 \label{eq:lambda}
\end{eqnarray}
The quantity $\lambda_\alpha$
in Eq. (\ref{eq:lambda}) is identified with 
 $M_\alpha^2-i\rho g_\alpha^2$ 
appearing in the usual Breit-Wigner formula.
Thus we define the physical mass $M_\alpha$ 
and the real physical coupling factor $g_\alpha$ 
($g_\alpha^2\equiv -{\rm Im}\ \lambda_\alpha /\rho$).
By a simple manipulation\footnote{ 
Equation (\ref{eq:Kresrep}) can be rewritten into
 the form 
$\rho {\cal T}^{\rm Res}=-{\rm Im}D/D$ (where
$D=(\bar m^2_{\bar\sigma}-s)(\bar m^2_{\bar f}-s)
-i\rho (\bar g^2_{\bar\sigma}(\bar m^2_{\bar f}-s)
+\bar g^2_{\bar f}(\bar m^2_{\bar \sigma}-s))$).
$D$ is factorized as 
$D=D_1D_2$, where $D_1=\lambda_\sigma-s$ and $D_2=\lambda_f-s$, 
and  $\rho {\cal T}^{\rm Res}$ is represented by  
$\rho {\cal T}^{\rm Res}
=-{\rm Im}(D_1D_2)/D_1D_2
=(-D_2{\rm Im}D_1-D_1{\rm Im}D_2
+2i{\rm Im}D_1{\rm Im}D_2)/D_1D_2$.
This is equivalent to Eq.(\ref{eq:IAderive}).
 }
 ${\cal T}^{\rm Res}$ is rewritten into the form
\begin{eqnarray}
{\cal T}^{\rm Res} &=& \frac{g_\sigma^2}{\lambda_\sigma-s}  
+\frac{g_f^2}{\lambda_f-s}+2i\rho 
\frac{g_\sigma^2}{\lambda_\sigma-s}\frac{g_f^2}{\lambda_f-s},
 \label{eq:IAderive}
\end{eqnarray}
where the $\lambda_\alpha$ and $g_\alpha$ are represented by 
$\bar m_{\bar\alpha}$, $\bar g_{\bar\alpha}$ 
and $\rho (s)$ (given in 
Eq. (\ref{eq:Gbar})), and, accordingly, are almost 
$s$ independent, except for the threshold region.
Thus, Eq. (\ref{eq:IAderive}) is understood to be just the form of
the scattering amplitude applied in IA method.
 
Similarly, ${\cal F}^{\rm Res}$ 
in the physical state representation
is given by
\begin{eqnarray}
{\cal F}^{\rm Res} &=& 
\frac{r_\sigma e^{i\theta_\sigma}}{\lambda_\sigma-s}  
+\frac{r_f e^{i\theta_f}}{\lambda_f-s},
 \label{eq:VMWderive}
\end{eqnarray}
where $r_\sigma e^{i\theta_\sigma}\equiv
\Sigma_\sigma F_\sigma$
and 
$r_f e^{i\theta_f}\equiv
\Sigma_f F_f$ ($\Sigma_\alpha 
(\equiv \bar g_{\bar\beta}u_{\bar\beta\alpha})$ is
the production coupling factor in the physical state representation, 
which is generally complex).
By using the equation
\begin{eqnarray}
{\cal F}^{\rm Res} &=& 
\frac{\bar r_{\bar\sigma}(\bar m_{\bar f}^2-s)
+\bar r_{\bar f}(\bar m_{\bar\sigma}^2-s)
}{(\lambda_\sigma-s)(\lambda_f-s)};
\ \ \ \bar r_{\bar\sigma}
\equiv\bar g_{\bar\sigma}\bar\xi_{\bar\sigma},
\ \ \bar r_{\bar f}\equiv\bar g_{\bar f}\bar\xi_{\bar f},
 \label{eq:Fres}
\end{eqnarray}
which is obtained from Eq. (\ref{eq:Presrep}),
 $r_\sigma ,r_f, \theta_\sigma$ and $\theta_f$
are given by
\begin{eqnarray}
r_\alpha e^{i\theta_\alpha} &=& 
[\bar r_{\bar\alpha}
(\bar m_{\bar\beta}^2-\lambda_\alpha )
+\bar r_{\bar\beta}
(\bar m_{\bar\alpha}^2-\lambda_\alpha )]/(\lambda_\beta
-\lambda_\alpha ),
\label{eq:con1}
\end{eqnarray}
where $(\alpha ,\beta )=(\sigma ,f)$ or $(f,\sigma )$.
As can be seen from Eq. (\ref{eq:con1}),
 $r_\alpha$ and $\theta_\alpha$ are almost 
$s$ independent, except for the threshold region.
Thus it is understood that
Eq. (\ref{eq:VMWderive}) is the same formula  
as that applied in VMW method.

In the VMW method, essentially the 
three new parameters,
$r_\sigma ,\ r_f$ and the relative phase 
$\theta (\equiv\theta_f -\theta_\sigma )$,
independent of the 
scattering process,
characterize the relevant production processes.
Presently they
are represented by the 
two production coupling constants, $\bar\xi_{\bar\sigma}$
and $\bar\xi_{\bar f}$.
Thus, among the three parameters in the VMW method there is
one constraint due to the FSI theorem.
 $r_\sigma$ and $r_f$, corresponding respectively to 
 $\bar r_{\bar\sigma}$ and $\bar r_{\bar f}$,
are regarded as free parameters.
On the other hand, $\theta (\equiv\theta_f-\theta_\sigma )$
is constrained by 
\begin{eqnarray}
Re^{i\theta} &=&
-\frac{(\bar m_{\bar f}^2-\lambda_f)
+\bar R(\bar m_{\bar\sigma}^2-\lambda_f)}{
(\bar m_{\bar f}^2-\lambda_\sigma )
+\bar R(\bar m_{\bar\sigma}^2-\lambda_\sigma )};\ \ \ \ 
(R\equiv r_f/r_\sigma ,
\ \ \bar R\equiv \bar r_{\bar f}/ \bar r_{\bar\sigma}),
\label{eq:constraint2}
\end{eqnarray}
which is obtained from Eq. (\ref{eq:con1}).

Next we consider 
the effect of the non-resonant background.
It can be
introduced consistently with 
the FSI theorem.
In the IA method the scattering
S matrix 
takes a multiplicative form of the resonant
and background parts as 
\begin{eqnarray}
S &=& S^{\rm Res}S^{\rm BG}=
\frac{1+i\rho{\cal K}^{\rm Res}}{1-i\rho{\cal K}^{\rm Res}}
\cdot\frac{1+i\rho{\cal K}^{\rm BG}}{1-i\rho{\cal K}^{\rm BG}}
\ \ .
\label{eq:SBG}
\end{eqnarray}
Correspondingly, the ${\cal T}$ matrix is represented by
the respective ${\cal K}$ matrices, ${\cal K}^{\rm Res}$ and
 ${\cal K}^{\rm BG}$.  ${\cal F}$ is obtained in a  
manner similar to Eq. (\ref{eq:Presrep}) as 
\begin{eqnarray}
{\cal T} &=& 
\frac{{\cal K}^{\rm Res}+{\cal K}^{\rm BG}}
{(1-i\rho{\cal K}^{\rm Res})(1-i\rho{\cal K}^{\rm BG})}
\rightarrow  
{\cal F} =
\frac{{\cal P}^{\rm Res}+{\cal P}^{\rm BG}}
{(1-i\rho{\cal K}^{\rm Res})(1-i\rho{\cal K}^{\rm BG})},
\label{eq:TBG}
\end{eqnarray}
where ${\cal P}^{\rm Res}$ and  ${\cal P}^{\rm BG}$ 
are the resonant and background production 
``${\cal K}$ matrix'', respectively.   
${\cal K}^{\rm BG}({\cal P}^{\rm BG})$ 
is equal to the background coupling factor  
$\bar g_{2\pi}(s)(\bar\xi_{2\pi}(s))$
in Eq. (\ref{eq:Lint}).\footnote{
The background coupling factor 
is generally 
$s$ dependent in the ``${\cal K}$ matrix"
representation.
For example,
in the case with the background of hard core type, 
$\bar g_{2\pi}(s)=-\frac{1}{\rho (s)}{\rm tan}\ p_1r_c.$
 }
This ${\cal F}$ automatically satisfies the FSI theorem.
It can be  rewritten as 
\begin{eqnarray}
{\cal F} &=& {\cal F}^{\rm Res}(1+i\rho{\cal T}_{\rm BG})
+{\cal F}^{\rm BG}(1+i\rho{\cal T}^{\rm Res}),
\label{eq:FBG2}\\
 &{\cal F}^{\rm BG}& = {\cal P}^{\rm BG}/(1-i\rho{\cal K}_{\rm BG})
\equiv f_{\rm BG}(s)e^{i\delta_{\rm BG}}.
\end{eqnarray}
Both the first and the second terms in Eq. (\ref{eq:FBG2})
have $\sigma$ and $f$ poles, and Eq. (\ref{eq:FBG2})
can be  rewritten as
\begin{eqnarray}
{\cal F} &=&
\frac{\bar r_{\bar\sigma}(s)(\bar m_{\bar f}-s)
+\bar r_{\bar f}(s)(\bar m_{\bar\sigma}-s)}{
(\lambda_\sigma -s)(\lambda_f -s)e^{-i\delta_{\rm BG}}}.
\label{eq:FBG3}
\end{eqnarray}
This has a similar form to 
Eq. (\ref{eq:Fres}), except for
the $s$ dependence of the production couplings,
given by
\begin{eqnarray}
\bar r_{\bar\sigma}(s) &=&
\bar r_{\bar\sigma}\ {\rm cos}\  \delta_{\rm BG}+f_{\rm BG}(s)
(\bar m_{\bar\sigma}^2-s),\ \ 
\bar r_{\bar f}(s) =
\bar r_{\bar f}\ {\rm cos}\  \delta_{\rm BG},
\label{eq:rs}
\end{eqnarray}
and the phase factor $e^{-i\delta_{\rm BG}}$.\footnote{
The overall phase factor 
$e^{-i\delta_{\rm BG}}$
appears only in the angular analysis
given in \S 2
through the scalar-tensor interfering term.
This factor has a weak $s$ dependence,
and its effect may be 
regarded as being
included in the phase parameters,
the $\theta_\alpha$, of the VMW method.
 }
In the case that  
the production coupling $\bar\xi_{2\pi}$ is so small
that $\bar r_{\bar\sigma}\gg f_{\rm BG}\bar m_{\bar\sigma}^2$,
$\bar r_{\bar\sigma}(s)$ and  $\bar r_{\bar f}(s)$ 
have weak $s$ dependences, and they 
are approximated with constants,
$\bar r_{\bar\sigma}(\bar m_{\bar\sigma}^2)$ and  
$\bar r_{\bar f}(\bar m_{\bar f}^2)$,
respectively. 
Then,  Eq. (\ref{eq:FBG3})
effectively  reduces to
 Eq. (\ref{eq:Fres}), 
and the VMW method
with a constrained phase parameter
is reproduced in this case with a non-resonant
background phase.  \\

\subsection{Physical Meaning of the 
``Universality of ${\cal T}_{\pi\pi}"$ and the VMW method}

\begin{figure}
 \epsfysize=15. cm
 \centerline{\epsffile{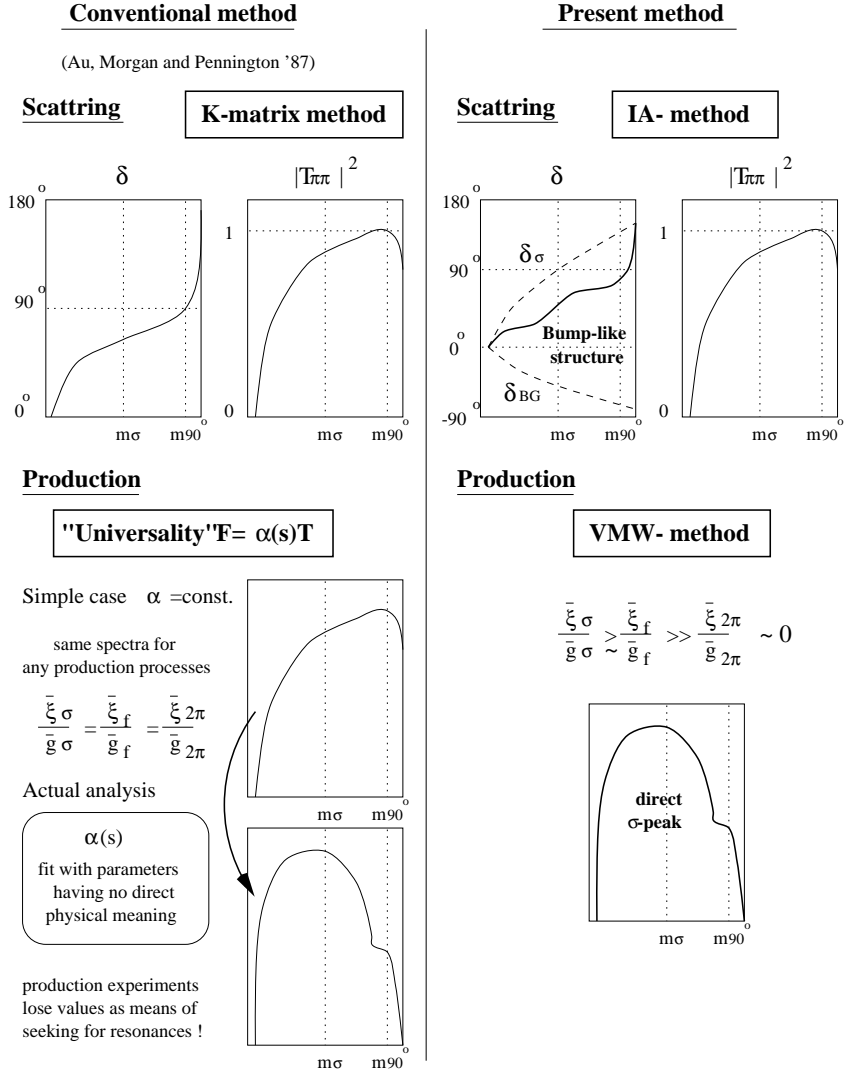}}
 \caption{Analyses by IA and  VMW methods
compared with the conventional analyses based on 
the ``Universality" of $\pi\pi$ scattering.}
\label{fig:compare}
\end{figure}

The methods of  
analyses we have used in studying scattering and production processes,
respectively, the IA and VMW methods,
are compared with those of 
the conventional analyses based on 
the ``universality" of $\pi\pi$ scattering 
pictorially in Fig. \ref{fig:compare}. 

The $\pi\pi$ scattering is largely affected by
the effect of the non-resonant repulsive background, 
and ${\cal T}$ cannot be described by the 
usual Breit-Wigner amplitudes alone 
with a non-derivative coupling.
The spectrum of ${\cal T}$ shows a very wide peak
around $\sqrt{s}\simeq 850$ MeV, at which value the phase
 $\delta_0^0$ passes through 90 degrees,
and then falls off rapidly, as shown in Fig. \ref{fig:compare}.
In contrast, 
the spectra of ${\cal F}$
in the $pp$ central collision and
the $J/\Psi\rightarrow\omega\pi\pi$-decay 
have peaks
at around $\sqrt{s}= m_\sigma$
 $(500\sim 600$ MeV).

In the conventional approach, with the universality 
relation ${\cal F}=\alpha {\cal T},$
 ${\cal T}$ is first analyzed and the 
phase shift $\delta$ around $\sqrt{s}=m_\sigma$
is interpreted as due to the background, 
instead of $\sigma$ contribution. 
Then ${\cal F}$ is analyzed with
 $\alpha (s)$ arbitrarily chosen 
in the polynomial form\footnote{
According to the relation ${\cal F}=\alpha {\cal T}$,
any production amplitude ${\cal F}$ vanishes at the same 
position as the zero position, $s=s_0^{\cal T}$, of ${\cal T}$,
and this is clearly incorrect. To avoid this problem, modified
forms, $\alpha (s)=\frac{1}{s-s_0^{\cal T}}
\sum_{n=0} \alpha_ns^n$\cite{rf:morg} or 
$\alpha (s)=\sum_{n=0} \bar\alpha_ns^n
+\frac{\lambda}{s-s_0^{\cal T}}$\cite{rf:MP2},
are used in the actual analyses. However, this operation is quite 
artificial and arbitrary, since 
we are free to choose any function which is zero at $s=s_0^{\cal T}$,
instead of $s-s_0^{\cal T}$,
to remove the zero of ${\cal T}$(${\cal K}$), 
as is seen from  the above two different forms
given by 
the original authors.
In our scheme the zero position of ${\cal F}$, $s=s_0^{\cal F}$,
is dependent on  both of the scattering and the production couplings 
of the relevant resonances, and is different from $s_0^{\cal T}$.
In this sense the above prescription for the ``common zeros'' problem
 is taken into account automatically.
(See further details in the criticism\cite{rf:penpen}
of our description and our reply\cite{rf:rep} to it.) 
 }
\begin{eqnarray}
\alpha (s) &=& \sum_{n=0} \alpha_ns^n.
\label{eq:poly}
\end{eqnarray}
In the most simple
case with $\alpha =$const, the universality relation
implies that
$\bar \xi_\sigma=\alpha \bar g_\sigma$,
$\bar \xi_f=\alpha \bar g_f$ and 
$\bar \xi_{2\pi}=\alpha \bar g_{2\pi}$;
that is, all the production couplings
are 
proportional to the corresponding $\pi\pi$ couplings,
and the spectra of ${\cal F}$ and ${\cal T}$
become the same.
Actually, they are different,
and the difference is fitted by  
$\alpha_n$.
The masses and widths of the resonances  
 are determined only from the 
$\pi\pi$ scattering, and 
the analyses of ${\cal F}$ on any production
process become 
 nothing but the determination of the
$\alpha_n$ for respective processes,
 which have no direct physical meaning. 
Thus all the production experiments 
lose their values in seeking new resonances. 

On the other hand, in the VMW method,
only the physically meaningful parameters are introduced.
The $\bar \xi_{\bar\sigma}$, $\bar \xi_{\bar f}$
and $\bar \xi_{2\pi}(s)$ are independent
parameters of the $\pi\pi$ scattering,
and
the difference between the spectra of
${\cal F}$ and ${\cal T}$ is explained
intuitively by supposing the relations among the   
coupling constants such as
\begin{eqnarray}
\frac{\bar\xi_{\bar\sigma}}{\bar g_{\bar\sigma}} & \gg &
\frac{\bar\xi_{2\pi}}{\bar g_{2\pi}},
\label{eq:ratio}
\end{eqnarray}
that is, 
the ratio of background effects to the $\sigma$-effects 
are weaker in 
the production processes
than in the scattering process.
Thus in this case  
the large low-energy peak structure in $|{\cal F}|^2$ 
shows directly the $\sigma$ existence.
In this situation
the properties of $\sigma$
can be obtained more precisely 
in the production processes
than in the scattering processes. It seems to us that 
this difference between the two methods reflects their
basic standpoints: In the ``universality" argument,
only the stable (pion) state forms 
the complete set of meson
states, while
 $\bar \sigma$ and $\bar f$, 
in addition to the pion,
are necessary as  bases of the complete set in the VMW method.

\section{Presence of the initial state phase and the VMW method}

\begin{figure}
 \epsfysize=15. cm
 \centerline{\epsffile{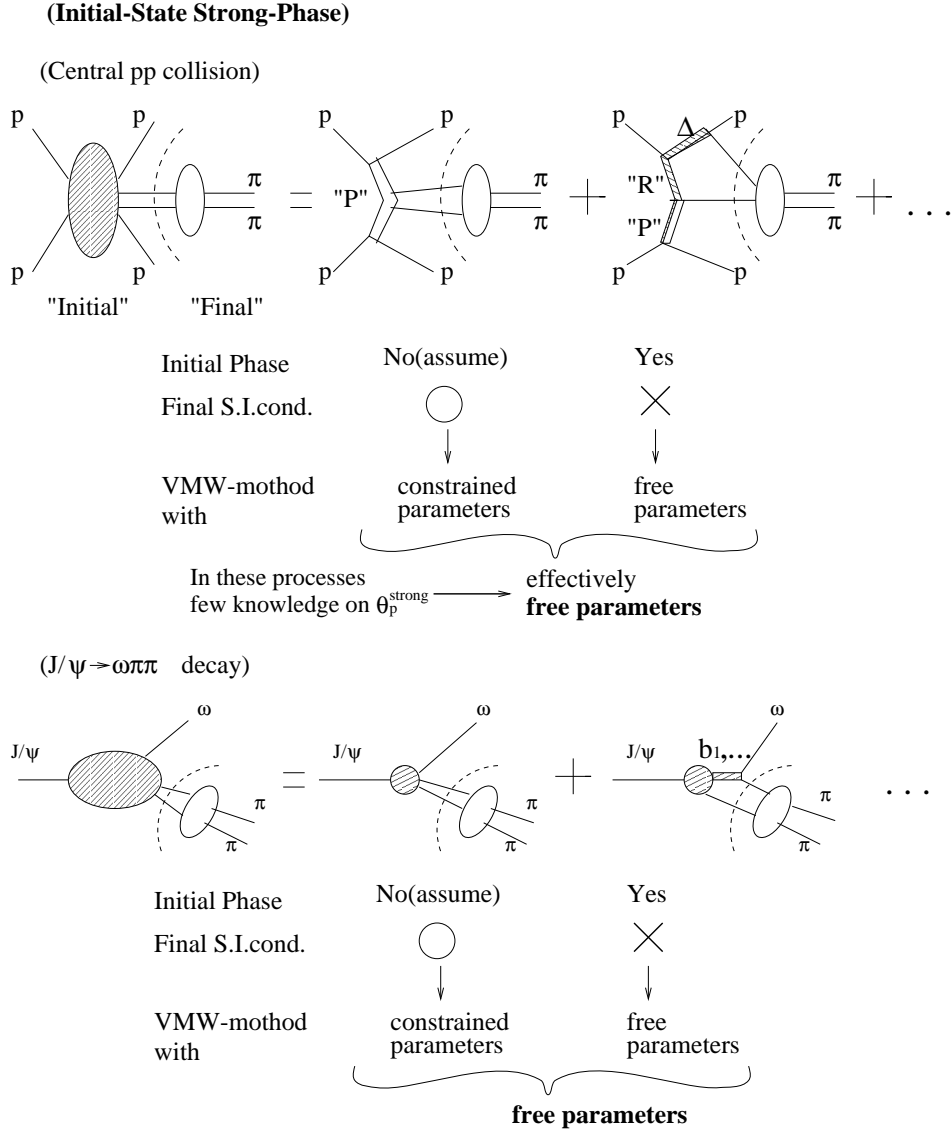}}
 \caption{Applicability of FSI theorem
to $pp$-central collision 
and $J/\Psi\rightarrow\omega\pi\pi$ decay. }
\label{fig:strong}
\end{figure}

In the previous section, we showed that  
the VMW method is an effective
method to determine the resonance properties 
from production processes,
although the parameters for  ${\cal F}$ have 
some constraint due to the FSI theorem.
Here it should be noted that 
the FSI theorem is only applicable
to the case in which the initial state has no strong phase.
We must carefully examine whether 
the initial state has a phase or not in actual cases.
We take the $pp$ central collision process, $pp\rightarrow pp\pi\pi$,
and the $J/\Psi\rightarrow\omega\pi\pi$-decay as examples. 
The situations are schematically shown in Fig. \ref{fig:strong}.

First, in the $pp$ central collision,
the main contribution to the relevant process is
usually considered to be due to 
 the double pomeron
exchange process, which is supposed to have no
strong phase.
However, it is known that this process 
is largely affected by the
$\Delta$-resonance diagram (described by the 
Breit-Wigner formula with finite width),
which causes the initial strong phase,
and the FSI condition may be violated.

A similar situation occurs also in 
the $J/\Psi\rightarrow\omega\pi\pi$-decay:
 this process is largely affected by the effect of 
$\omega\pi$-resonances, such as $b_1(1235)$ and its 
excited states, which supply the initial strong phase.

This type of initial strong phase 
generally exists in all processes 
under the effect of strong interactions,\footnote{
However, in weak decays, such as $K\rightarrow 2\pi$ and
$K_{l4}$ decays, the FSI condition is exactly satisfied,
and the analysis using the VMW method with
free parameters is not applicable
to these processes.}
which can effectively 
be introduced in the VMW method by
substitution of 
\begin{eqnarray}
\bar r_{\bar\alpha} &\rightarrow & \bar r_{\bar\alpha}
e^{i\bar\theta_{\bar\alpha}^{\rm str}}.
\label{eq:stphase}
\end{eqnarray}
We have little knowledge of these initial 
phases, and we are forced to treat the parameters 
in the VMW method as being effectively free.

\section{Concluding remarks}

In this paper the relation between scattering and production 
amplitudes was investigated from the general viewpoint 
of the unitarity and the applicability of the FSI theorem, 
by using a simple field-theoretical model. The methods used in our 
phenomenological analyses of the $\pi\pi$-phase shift and 
the production processes, the IA method and the VMW method, 
respectively, were derived directly 
in the physical state representation of scattering and production 
amplitudes. The relative phase parameters $\theta$ in the VMW method  
are constrained by the FSI theorem in the case it is applicable.
However, in general production processes under the effect of 
strong interactions, the initial states have unknown strong phases,
and correspondingly the $\theta$ in VMW method are treated as being 
effectively free. 

Furthermore, we have checked carefully the physical meaning of
the ``universality'' argument, and 
have argued that the conventional analyses 
following it seem to be only  
parameter-fitting and meaningless in seeking 
new resonances, while the VMW method is an effective method
applicable to general production processes 
under the effect of strong interactions,
in determining 
the existence and properties of new resonances.

For many years, some 
experimental 
facts\cite{rf:WA,rf:CB,rf:Ak,rf:hiro}
 suggesting $\sigma$ existence, other than mentioned in the Introduction,
 obtained in various production processes,  
had been persuaded to be interpreted without the 
$\sigma$ meson  by invoking the 
``universality'' of $\pi\pi$-scattering amplitude.  
However, as is shown here, we may conclude that 
the conventinal argument of the ``universality''
is incorrect. Accordingly, these  
production experiments should be 
re-analyzed through the VMW method by taking into account the 
possible effects of $\sigma$ existence.
Especially in this connection the negative estimation\cite{rf:pennington}
on the results\cite{rf:GAMS,rf:taku1,rf:taku2}
obtained from the GAMS experiment (, which was mentioned in the 
Introduction,) should be corrected.

\section*{Acknowledgements}
We would like to express our sincere gratitude to 
Professor K. Takamatsu
and Professor T. Tsuru for valuable comments and continual encouragement.

\end{document}